\documentclass[epj]{svjour}
\usepackage{graphics}
\begin{document}
\title{Mass spectra of four quark states in hidden charm sector }
\subtitle{}
\author{Smruti Patel, Manan Shah \and P C Vinodkumar
% \thanks is optional - remove next line if not needed
%\thanks{{e-mail:} fizix.smriti@gmail.com}
\thanks{{e-mail:} p.c.vinodkumar@gmail.com}
}                     % Do not remove
%
%\offprints{}          % Insert a name or remove this line
%
\institute{Department of Physics, Sardar Patel University,Vallabh Vidyanagar, INDIA. }
\date{Received: date / Revised version: date}
% The correct dates will be entered by Springer
%
\abstract{
Masses of the low lying four quark states in the hidden charm sector ($cq\bar c \bar q; q\in u,d$) are calculated within the framework of a non-relativistic quark model. The four body system is considered as two two-body systems such as diquark-antidiquark  ($Qq-\bar Q \bar q$) and  quark antiquark-quark antiquark ($Q\bar q -\bar Qq$) molecular-like four quark states. Here, Cornell type potential has been used for describing the two body interactions among $Q-q$, $\bar Q-\bar q$, $Q-\bar q$, $Qq-\bar Q \bar q$ and $Q\bar q-\bar Qq$, with appropriate string tensions. Our present analysis suggests the following exotic states, $X(3823)$, $Z_c(3900)$, $X(3915)$, $Z_c(4025)$, $\psi(4040)$, $Z_1(4050)$ and $X(4160)$ as $Q\bar q-\bar Qq$ molecular-like four quark states while $Z_c(3885)$, $X(3940)$ and $Y(4140)$ as the diquark-antidiquark  four quark states. We have been
able to assign the $J^{PC}$ values for many of the recently observed exotic states according to their structure. Apart from this, we have identified the charged state $Z(4430)$ recently confirmed by LHCb as
the first radial excitation of $Zc(3885)$ with G=+1 and $Y(4360)$ state as the first radial excitation of $Y(4008)$ with $G=-1$ and the
state $\psi(4415)$ as the first radial excitation of the $\psi(4040)$ state.
\PACS{
      {PACS-key}{Exotic, tetraquark}
     } % end of PACS codes
} %end of abstract
\maketitle

\section{Introduction} \label{intro}
Over the past decade, the family of exotic states has become more and more abundant due to the  development at the experimental front. It is a topic of current interest full of opportunities and challenges for theorists as well as experimentalists to reveal the internal mechanisms originating from these novel and complicated states. Many exotic states in the charm sector with c$\bar c$ content have been discovered by Belle \cite{liu} and BESIII \cite{M Abl} and  others which provide challenges to theorists studying  hadron spectroscopy. With the experimental progress, theorists have paid more attention to these observations by proposing different explanations.  Due to the asymptotic property of QCD, study of the hadron physics have to concern about the nonperturbative effect which is difficult in quantum field theory. It has been realized early on that quark models and QCD sustain a much richer pattern of different multi-quark and/or color network configurations, beyond the "non-exotic" standard $q\bar q$ mesons and qqq baryons. There are growing evidences for the existence of exotic mesons containing both heavy and light quark-antiquark pairs i.e. $c\bar cq\bar q$. In the past few years, the experimental observations of large numbers of X, Y and Z states have stimulated the study of exotic states greatly as they have induced a pre Gell- Mann like situation in our knowledge of the hadron spectroscopy.

Definite conclusions have not yet been reached about the internal structure of newly observed four quark systems. Models to accommodate the exotic states have been proposed over the last decades. Different attempts have been made for the interpretation of the internal structure of the exotic hadronic states of four quark system. These four quark state can in principle be composed of diquark-antidiquark  (a tetraquark), a loosely bound state of two mesons (a molecular state), glueballs or $q\bar q$ pair with gluons (hybrids). Here, we confine ourselves to the study of four quark exotic states containing hidden charm($c, \bar c$) with light flavor($u, d$) combinations. Study of such structures is important from the point of view of understanding interaction among hadrons at different energy scales related to their formation of bound states as well as their decay processes. These interactions provide useful information to study fundamental problems of QCD such as color confinement. The proposal to revisit the multiquark picture using diquarks has been raised by Jaffe and Wilczek \cite{JW}. The first papers suggesting the existence of tetraquark configurations were given by \cite{Jaffe,R}, within the MIT bag model with color spin interaction. In the beginning, light flavor tetraquark states were predicted. The possible exotic structure for the interpretations of the light scalar mesons was discussed in diquark-antidiquark  picture in Refs \cite{Jaffe,hooft,Lmaiani,black}.  In Refs \cite{hooft,Lmaiani}, authors  presented  decay of light scalar meson in the diquark -antidiquark picture and discussed a mixing of light scalars with that of the positive parity $q\bar q$ states. Later on, Weinstein and Isgur \cite{W,Isgur} extended this tetraquark picture into a variety of quark models. This means that tetraquarks with heavy quarks can also exist. In the past thirty five  years, theorists have been studying whether two charmed mesons can be bound into a molecular state as the presence of the heavy quarks lowers the kinetic energy while the interaction between two light quarks could still provide a strong enough attraction. Voloshin and Okun studied the interaction between a pair of charmed mesons and proposed the possibilities of the molecular states involving charmed quarks \cite{Voloshin}. Several attempts have been made to describe the interaction between two mesons in the di-mesonic molecular state based on potential models \cite{Youchang,Wong,Rui}. The states $Z_c(3900)$ reported by BELLE and BES-III \cite{liu,M Abl} and $Z(4430)$ discovered earlier by BELLE\cite{Choi} and confirmed recently by LHCb\cite{RA} are of recent interest in the study of exotic hadrons. These exotic states are interpreted as strong candidates for the tetraquark states\cite{Dias,LM}. Another interesting possible interpretation of the $Z_c(3900)$ as a molecular like state of 1/($\sqrt{2}$)($D\bar D^{*+}D^*\bar D$) resulting from the binding of two charmed mesons is proposed by Hong et al\cite{Hong}. Thus, its interpretation as a tetraquark state or a di-mesonic molecular state remains unresolved. We make an attempt to predict masses of heavy tetraquarks in the framework of a non-relativistic quark model based on the potential approach. Here we consider the four-body problem in terms of the subsequent more simple three two-body systems.

 %The electromagnetic transitions are also  calculated in this scheme because these radiative transitions can probe the internal charge structure of mesons and hence are very useful in determining the meson structure.

\section{Theoretical Framework}
In this paper we shall take a different path and investigate different ways in which the experimental data can be reproduced. There are many methods to estimate the mass of a hadron, among which phenomenological potential model is a fairly reliable one.

Non-relativistic interaction potential we have used here is the Cornell potential consists of a central term V(r) which is being just a  sum of the Coulomb(vector) and linear confining(scalar) parts given by

\begin{equation}\label{eq:9}
V(r)=V_V+V_S=k_s\frac{\alpha_s}{r}+\sigma r
\end{equation}
\begin{eqnarray}
k_s&=&-4/3 \ \ for \ \ q\bar q\ \ \ \nonumber \\
   &=&-2/3 \ \ for \ \ qq \ \ or \ \ \bar q\bar q
\end{eqnarray}

Different degenerate exotic states can be calculated by including spin-dependent part of the usual one gluon exchange potential. The potential description extended to spin dependent interactions results in three types of interaction terms such as the spin-spin , the spin-orbit and the tensor part that are to be added to the discussed leading non-relativistic description. Accordingly, the spin-dependent part $V_{SD}$  is given by

\begin{eqnarray}\label{eq:10}
V_{SD}&=&V_{SS}\left[\frac{1}{2}(S(S+1)-S_1(S_1+1)-S_2(S_2+1))\right] \nonumber \\ && +V_{LS}\left[\frac{1}{2}(J(J+1)-S(S+1)-L(L+1))\right] \nonumber \\ && +V_{T}\left[12\left(\frac{(S_1.r)(S_2.r)}{r^2}-\frac{1}{3}(S_1.S_2)\right)\right]
\end{eqnarray}\\

The spin-orbit term containing $V_{LS}$ and tensor term containing $V_T$ describe the fine structure of the states, while the spin-spin term containing $V_{SS}$ proportional to 2$S_1.S_2$ gives the hyperfine splitting. The coefficient of these spin-dependent terms of Eq.(\ref{eq:10}) can be written in terms of the vector and scalar parts of static potential V(r) as

\begin{equation}\label{eq:11}
V^{ij}_{LS}(r)=\frac{1}{2M_iM_jr}+\left[ 3\frac{dV_V}{dr}-\frac{dV_S}{dr}\right]
\end{equation}

\begin{equation}\label{eq:12}
V^{ij}_T(r)=\frac{1}{6M_iM_jr}+\left[ 3\frac{d^2V_V}{dr^2}-\frac{1}{r}\frac{dV_S}{dr}\right]
\end{equation}

\begin{equation}\label{eq:13}
V^{ij}_{SS}(r)=\frac{1}{3M_iM_jr}\nabla^2 V_V=\frac{16\pi\alpha_s}{9M_iM_j}\delta^3(r)
\end{equation}

Where $M_i$, $M_j$ corresponds to the masses and r is relative co-ordinate of the two body system under consideration.\\
Our main aim is to interpret the four quark state structure in two different schemes: (1) Clusters of diquark-antidiquark  like structure and (2) Clusters of quark antiquark-quark antiquark like structure. In this case, one uses the fact that the motion of the quarks that comprise the exotic state is nonrelativistic to assume that they move in a static potential, much like
nonrelativistic models of the hydrogen atom. In both the pictures, we have treated the four particle system as
two-two body systems interacting through effective potential of the same form of the two body interaction potential discussed above but with different
interaction strengths, $k_s$ and $\sigma$ of Eq. (\ref{eq:9}). \\ \\
The model parameters including the constituent quark masses ($m_{u/d}$, $m_c$) and string tension ($\sigma$) are chosen to get the ground state masses of experimentally observed $X(3823)$, $Z_c(3823)$ and $Z_c(3885)$ exotic states. The fitted model parameters are listed in the Table 1 along with computed ground state radii of $Qq$, $Q\bar q$, $Qq-\bar Q\bar q$ and $Q\bar q-\bar Qq$ systems.
From the radii listed in the Table 1 we can see that though both $Qq$, $Q\bar q$ systems having same string tension the size of $Qq$ system is larger than the $Q\bar q$ system as the color coulomb part becomes less attractive in the case of $Qq$ system compared to $Q\bar q$ system. Also the size of $Qq-\bar Q\bar q$ is slightly smaller than that of $Q\bar q-\bar Qq$ system indicative of the fact that the inter-cluster residual color interaction in the case of $Q\bar q-\bar Qq$ is weaker relative to $Qq-\bar Q\bar q$.
 The two body Schr\"{o}dinger equation is numerically solved using Range-Kutta method of mathematica
notebook \cite{Lucha,BKP}.

%%%%%%%%%%%%%%%%%%%%%%%%%%%%%%%%%%%%%%%%%%%%%%%%%%%%%%%%%%%%%%%%%%%%%%%%%%%%%%%%%%%%%%%%%%%%%%%%%%%%%%%%%%%%%%%%%%%%%%%%%

\begin{table*}
\begin{center}
\caption{Fitted model parameters (quark mass and string tension) along with computed ground state rms radii.} \label{tab1}
\begin{tabular}{cc|ccc}
\hline\hline
&Quark masses& System               & String tension $\sigma$  &          $\langle r^2\rangle^{1/2} $ \\
& &  & ($ GeV^2$)&(fm) \\
\hline
& $m_u=m_d=0.323 GeV$& $Qq$                 &         0.015                  &             1.539 \\
& $m_c=1.486 GeV$& $Qq-\bar Q\bar q$    &         0.030                  &           0.513 \\
&& $Q\bar q$            &         0.015                  &           1.393 \\
&& $Q\bar q-\bar Q q$   &         0.018                 &           0.530 \\
\hline\hline

\end{tabular}
%\begin{center}
% $m_{u/d}=0.323 GeV$, $m_c=1.486 GeV$
%\end{center}
\end{center}
\end{table*}
%%%%%%%%%%%%%%%%%%%%%%%%%%%%%%%%%%%%%%%%%%%%%%%%%%%%%%%%%%%%%%%%%%%%%
\begin{table*}
\begin{center}
\caption{Mass spectra of four quark state with $Qq-\bar Q\bar q$ structure for $L_d$=0 and $L_{\bar d}$=0 (in GeV).} \label{tab2}
\begin{tabular}{cccccccccccccc}
\hline\hline\\

$S_d$	&	$L_d$	&	$S_{\bar d}$	&	$L_{\bar d}$	&	$J_d$	&	$J_{\bar d}$	&	J	&	$J^{PC}$ 	 &	$^{2S+1}X_J$	&$M_{cw}$&	 $\langle V_{SS}\rangle$	&	$\langle V_{LS}\rangle$	&	 $\langle V_{T}\rangle$	 &	Mass	 \\
\hline\\
0	&	0	&	0	&	0	&	0	&	0	&	0	&	$0^{++}$	&	$^1S_{0}$	&	3.906	&	0	&	0	 &	0	&	3.906	\\ \\
1	&	0	&	0	&	0	&	1	&	0	&	1	&	$1^{+-}$	&	$^3S_{1}$	&	3.910	&	0	&	0	 &	0	&	3.910	\\ \\
1	&	0	&	1	&	0	&	1	&	1	&	0	&	$0^{++}$	&	$^1S_{0}$	&	3.914	&	-0.0650	&	 0	&	0	&	3.849	\\
	&		&		&		&		&		&	1	&	$1^{+-}$	&	$^3S_{1}$	&		&	-0.0325	&		 &		&	3.882	\\
	&		&		&		&		&		&	2	&	$2^{++}$	&	$^5S_{2}$   &		&	0.0325	&		 &		&	3.946	\\

\hline\hline

\end{tabular}
\begin{center}
$M_{cw}$-center of weight mass
\end{center}
\end{center}

\end{table*}
%%%%%%%%%%%%%%%%%%%%%%%%%%%%%%%%%%%%%%%%%%%%%%%%%%%%%%%%%%%%%%%%%%%%%%%%%%%%%%%%%%%%%%%%%%%%%%%%%%%%%%%%%%%%%%%%
\begin{table*}
\begin{center}
\caption{Mass spectra of four quark state with $Qq-\bar Q\bar q$ structure for $L_d$=1 and $L_{\bar d}$=0 (in GeV).} \label{tab3}
\begin{tabular}{cccccccccccccc}
\hline\hline\\

$S_d$	&	$L_d$	&	$S_{\bar d}$	&	$L_{\bar d}$&	$J_d$	&	$J_{\bar d}$	&	J	&	$J^{PC}$ 	 &	$^{2S+1}X_J$	&	$M_{cw}$	&	 $\langle V_{SS}\rangle$	&	$\langle V_{LS}\rangle$	&	 $\langle V_{T}\rangle$&	Mass	\\
\hline\\
0	&	1	&	0	&	0	&	1	&	0	&	1	&	$1^{- -}$	&	$^1P_1$	&	4.156	&	0	&	0	&	 0.0082	&	4.164	\\ \\
1	&	1	&	0	&	0	&	0	&	0	&	0	&	$0^{- +}$	&	$^3P_0$	&		&		&	-0.0035	 &	 -0.0165	&	4.136	\\
	&		&		&		&	1	&		&	1	&	$1^{- +}$	&	$^3P_1$	&	4.156	&	0	&	-0.0017	 &	 0.0041	&	4.159	\\
	&		&		&		&	2	&		&	2	&	$2^{- +}$	&	$^3P_2$	&		&		&	0.0017	&	 -0.0041	&	4.154	\\ \\
1	&	1	&	1	&	0	&	0	&	1	&	1	&	$1^{- -}$	&	$^1P_1$	&	4.160	&	-0.0017	&	0	 &	0	&	4.145	\\ \\
	&		&		&		&	1	&	1	&	0	&	$0^{- +}$	&	$^3P_0$	&		&		&	-0.0035	 &	 -0.027	&	4.128	\\
	&		&		&		&		&		&	1	&	$1^{- +}$	&	$^3P_1$	&4.160		&	-0.0008	&	 -0.0017	&	 -0.0688	&	4.151	\\
	&		&		&		&		&		&	2	&	$2^{- +}$	&	$^3p_2$	&		&		&	0.0017	&	 -0.0151	&	4.146	\\ \\
	&		&		&		&	2	&	1	&	1	&	$1^{- -}$	&	$^5P_1$	&		&		&	-0.0052	 &	 -0.0426	&	4.113	\\
	&		&		&		&		&		&	2	&	$2^{- -}$	&	$^5P_2$	&	4.160	&	0.00088	&	 -0.0017	&	 0.0151	&	4.174	\\
	&		&		&		&		&		&	3	&	$3^{- -}$	&	$^5P_3$	&		&		&	0.0035	&	 -0.022	&	4.142	\\

\hline\hline

\end{tabular}

\end{center}

\end{table*}

%%%%%%%%%%%%%%%%%%%%%%%%%%%%%%%%%%%%%%%%%%%%%%%%%%%%%%%%%%%%%%%%%%%%%%%%%%%%%%%%%%%%%%%%%%%%%%%%%%%%%%%%%%%%%%%

%%%%%%%%%%%%%%%%%%%%%%%%%%%%%%%%%%%%%%%%%%%%%%%%%%%%%%%%%%%%%%%%%%%%%%%%%%%%%%%%%%%%%%%%%%%%%%%%%%%%%%%%%%%%%%
\begin{table*}
\begin{center}
\caption{Mass spectra of four quark state with $Q\bar q-\bar Qq$ molecular-like structure for $L_1$=0 and $L_2$=0 (in GeV).} \label{tab4}
\begin{tabular}{cccccccccccccccc}
\hline\hline\\

$S_1$	&	$L_1$	&	$S_2$	&	$L_2$&	$J_1$	&	$J_2$	&	$J_{12}$	&	$L_{12}$	&	J	&	 $J^{PC}$  	&	$^{2S+1}X_J$	&	 $M_{cw}$	&	 $\langle V_{SS}\rangle$	&	$\langle V_{LS}\rangle$	&	 $\langle V_{T}\rangle$	&	Mass	\\
\hline\\
0	&	0	&	0	&	0	&	0	&	0	&	0	&	0	&	0	&	$0^{-+}$	&	$^1S_0$	&	3.817	 &	 0	&	0	&	0	&	3.817	\\
	&		&		&		&		&		&		&	1	&	1	&	$1^{+-}$	&	$^1P_1$	&	3.929	 &	 0	&	0	&	0.0057	&	3.935	\\
	&		&		&		&		&		&		&	2	&	2	&	$2^{-+}$	&	$^1D_2$	&	3.999	 &	 0	&	0	&	0.00094	&	4.000	\\
\hline\\
1	&	0	&	0	&	0	&	1	&	0	&	1	&	0	&	1	&	$1^{+-}$	&	$^3S_1$		&	 3.817	 &	0	&	0	&	0	&	3.817	\\
	&		&		&		&		&		&		&		&		&		&		&		&		&		&		 &		\\
	&		&		&		&		&		&		&		&	0	&	$0^{++}$	&	$^3P_0$	&		 &	 0	 &	-0.0022	&	-0.0011	&	3.915	\\
	&		&		&		&		&		&		&	1	&	1	&	$1^{++}$	&	$^3P_1$	&	3.929	 &	 0	 &	-0.0011	&	0.0028	&	3.931	 \\
	&		&		&		&		&		&		&		&	2	&	$2^{++}$	&	$^3P_2$	&		&	 0	 &	0.0011	&	-0.0028	&	3.927	\\
	&		&		&		&		&		&		&		&		&		&		&		&		&		&		 &		\\
	&		&		&		&		&		&		&		&	1	&	$1^{--}$	&	$^3D_1$	&		 &	 0	 &	-0.0002	&	-0.0014	&	3.997	\\
	&		&		&		&		&		&		&	2	&	2	&	$2^{--}$	&	$^3D_2$	&	3.997	 &	 0	 &	-0.00006	&	0.0008	&	 4.000	\\
	&		&		&		&		&		&		&		&	3	&	$3^{--}$	&	$^3D_3$	&		&	 0	 &	0.00012	&	-0.0006	&	3.999	\\
\hline\\

1	&	0	&	1	&	0	&	1	&	1	&	0	&		&	0	&	$0^{-+}$	&	$^1S_0$	&		&	 -0.0069	&	0	&	0	&	3.747	\\
	&		&		&		&		&		&	1	&	0	&	1	&	$1^{--}$	&	$^3S_1$	&	3.817	 &	 -0.0034	&	0	&	0	&	3.782	 \\
	&		&		&		&		&		&	2	&		&	2	&	$2^{-+}$	&	$^5S_2$	&		&	 0.0034	&	0	&	0	&	3.853	\\
	&		&		&		&		&		&		&		&		&		&		&		&		&		&		 &		\\
	&		&		&		&		&		&	0	&	1	&	1	&	$1^{+-}$	&	$^1P_1$	&	3.929	 &	 -0.0010	&	0	&	-0.0096	&	 3.919	\\
	&		&		&		&		&		&		&		&		&		&		&		&		&		&		 &		\\
	&		&		&		&		&		&		&		&	0	&	$0^{++}$	&	$^3P_0$	&		&		 &	-0.0022	&	-0.019	&	3.907	\\
	&		&		&		&		&		&	1	&	1	&	1	&	$1^{++}$	&	$^3P_1$	&	3.929	 &	 -0.0005	&	-0.0011	&	-0.0048	&	 3.923	 \\
	&		&		&		&		&		&		&		&	2	&	$2^{++}$	&	$^3P_2$	&		&		 &	0.0011	&	-0.010	&	3.919	\\
	&		&		&		&		&		&		&		&		&		&		&		&		&		&		 &		\\
	&		&		&		&		&		&		&		&	1	&	$1^{+-}$	&	$^5P_1$	&		&		 &	-0.0033	&	-0.029	&	3.897	\\
	&		&		&		&		&		&	2	&	1	&	2	&	$2^{+-}$	&	$^5P_2$&	3.993	 &	 0.0005	&	-0.0011 &	-0.010	&	 3.939	\\
	&		&		&		&		&		&		&		&	3	&	$3^{+-}$	&	$^5P_3$&		&		 &	0.0022	&	-0.015	&	3.917	\\

\hline\hline

\end{tabular}

\end{center}

\end{table*}

%%%%%%%%%%%%%%%%%%%%%%%%%%%%%%%%%%%%%%%%%%%%%%%%%%%%%%%%%%%%%%%%%%%%%%%%%%%%%%%%%%%%%%%%%%%%%%%%%%%%%%%%%%%%%%%%

\begin{table*}
\begin{center}
\caption{Mass spectra of four quark state with $Q\bar q-\bar Qq$ molecular-like structure for $L_1$=1 and $L_2$=0 (in GeV).} \label{tab5}
\begin{tabular}{cccccccccccccccc}
\hline\hline\\ \\

$S_1$	&	$L_1$	&	$S_2$	&	$L_2$&	$J_1$	&	$J_2$	&	$J_{12}$	&	$L_{12}$	&	J	&	 $J^{PC}$  	&	$^{2S+1}X_J$	&	 $M_{cw}$	&	 $\langle V_{SS}\rangle$	&	$\langle V_{LS}\rangle$	&	 $\langle V_{T}\rangle$&	Mass	\\
\hline\\
0	&	1	&	0	&	0	&	1	&	0	&	1	&	0	&	1	&	$1^{--}$	&	$^3S_1$	&	3.927	 &	 0	&	0	&	0	&	3.927	\\
	&		&		&		&		&		&		&		&		&		&		&		&		&		&		 &\\		
	&		&		&		&		&		&		&		&	0	&	$0^{++}$	&	$^3P_0$	&		&		 &	-0.0022	&	-0.011	&	4.023	\\
	&		&		&		&		&		&		&	1	&	1	&	$1^{++}$	&	$^3P_1$	&	4.036	 &	 0	&	-0.0011	&	0.0028	&	4.038	 \\
	&		&		&		&		&		&		&		&	2	&	$2^{++}$	&	$^3P_2$	&		&		 &	0.0011	&	-0.0028	&	4.035	\\
	&		&		&		&		&		&		&		&		&		&		&		&		&		&		 &		\\
	&		&		&		&		&		&		&		&	1	&	$1^{--}$	&	$^3D_1$	&		&		 &	-0.0002	&	-0.0013	&	4.104	\\
	&		&		&		&		&		&		&	2	&	2	&	$2^{--}$	&	$^3D_2$	&	4.106	 &	 0	&	-0.00006	&	0.0007	&	 4.107	\\
	&		&		&		&		&		&		&		&	3	&	$3^{--}$	&	$^3D_3$		&		 &		 &	0.00013	&	-0.006	&	4.106	 \\
\hline\\
	&		&		&		&		&		&		&		&		&		&		&		&		&		&		 &		\\
1	&	1	&	1	&	0	&	0	&	1	&	1	&	0	&	1	&	$1^{--}$ &	$^3S_1$	&	3.927	&	 0	 &	0	&	0	&	3.927	\\
	&		&		&		&		&		&		&		&		&		&		&		&		&		&		 &		\\
	&		&		&		&		&		&	0	&		&	0	&	$0^{-+}$	&	$^1S_0$	&		&	 -0.069	&	0	&	0	&	3.858	\\
	&		&		&		&	1	&	1	&	1	&	0	&	1	&	$1^{--}$	&	$^3S_1$	&	3.927	 &	 -0.034	&	0	&	0	&	3.892	\\
	&		&		&		&		&		&	2	&		&	2	&	$2^{-+}$	&	$^5S_2$	&		&	 0.034	 &	0	&	0	&	3.962	\\
	&		&		&		&		&		&		&		&		&		&		&		&		&		&		 &		\\
	&		&		&		&		&		&	1	&		&	1	&	$1^{--}$	&	$^3S_1$	&		&	 -0.0910	&	0	&	0	&	3.823	\\
	&		&		&		&	2	&	1	&	2	&	0	&	2	&	$2^{-+}$	&	$^5S_2$	&	3.927	 &	 -0.034	&	0	&	0	&	3.892	\\
	&		&		&		&		&		&	3	&		&	3	&	$3^{--}$		&	$^7S_3$	&		 &	 0.069	&	0	&	0	&	3.996	\\
	&		&		&		&		&		&		&		&		&		&		&		&		&		&		 &		\\
	&		&		&		&	0	&	1	&	1	&	1	&	0	&	$0^{++}$	&	$^3P_0$	&		&	 0	 &	-0.0022	&	-0.011	&	4.023	\\
	&		&		&		&		&		&		&		&	1	&	$1^{++}$	&	$^3P_1$	&	4.036	 &	 0	&	-0.0011	&	0.0028	&	4.038	 \\
	&		&		&		&		&		&		&		&	2	&	$2^{++}$	&	$^3P_2$	&		&	 0	 &	0.0011	&	-0.0028	&	4.035	\\
	&		&		&		&		&		&		&		&		&		&		&		&		&		&		 &		\\
	&		&		&		&	1	&	1	&	0	&	1	&	1	&	$1^{+-}$		&	$^1P_1$	&	 4.036	 &	-0.0009	&	0	&	-0.0094	&	 4.026	\\
	&		&		&		&		&		&		&		&		&		&		&		&		&		&		 &		\\
	&		&		&		&		&		&		&		&	0	&	$0^{++}$		&	$^3P_0$	&		 &		 &	-0.0022	&	-0.018	&	4.015	 \\
	&		&		&		&		&		&	1	&	1	&	1	&	$1^{++}$		&	$^3P_1$	&	 4.036	 &	-0.0004	&	-0.0011	&	-0.0047	&	 4.030	 \\
	&		&		&		&		&		&		&		&	2	&	$2^{++}$		&	$^3P_2$	&		 &		 &	0.0011	&	-0.010	&	4.027	 \\
	&		&		&		&		&		&		&		&		&		&		&		&		&		&		 &		\\
	&		&		&		&		&		&		&		&	1	&	$1^{+-}$	&	$^5P_1$	&		&		 &	-0.0033	&	-0.029	&	4.004	\\
	&		&		&		&		&		&	2	&	1	&	2	&	$2^{+-}$	&	$^5P_2$&	4.036	 &	 0.0004	&	-0.0011	&	0.010	&	 4.046	\\
	&		&		&		&		&		&		&		&	3	&	$3^{+-}$	&	$^5P_3$	&		&		 &	0.0022	&	-0.015	&	4.024	\\
	&		&		&		&		&		&		&		&		&		&		&		&		&		&		 &		\\
	&		&		&		&	2	&	1	&	1	&	1	&	0	&	$0^{++}$	&	$^3P_0$	&		&		 &	-0.0022	&	-0.034	&	3.999	\\
	&		&		&		&		&		&		&		&	1	&	$1^{++}$	&	$^3P_1$	&	4.036	 &	 -0.0014	&	-0.0011	&	-0.019	&	 4.015	 \\
	&		&		&		&		&		&		&		&	2	&	$2^{++}$	&	$^3P_2$	&		&		 &	0.0011	&	-0.025	&	4.010	\\
	&		&		&		&		&		&		&		&		&		&		&		&		&		&		 &		\\
	&		&		&		&		&		&	2	&		&	1	&	$1^{--}$	&	$^5P_1$	&		&		 &	-0.0033	&	-0.044	&	3.988	\\
	&		&		&		&		&		&		&		&	2	&	$2^{--}$	&	$^5P_2$	&	4.036	 &	 -0.00048	&	-0.0011	&	-0.047&	 4.030	\\
	&		&		&		&		&		&		&		&	3	&	$3^{--}$	&	$^5P_3$	&		&		 &	0.0022	&	-0.030	&	4.008	\\
	&		&		&		&		&		&		&		&		&		&		&		&		&		&		 &		\\
	&		&		&		&		&		&	3	&		&	2	&	$2^{++}$	&	$^7P_2$	&		&		 &	-0.0044	&	-0.058	&	3.974	\\
	&		&		&		&		&		&		&		&	3	&	$3^{++}$	&	$^7P_3$		& 4.036		 &	 0.0009	&	-0.0011	&	0.018	&	 4.054	 \\
	&		&		&		&		&		&		&		&	4	&	$4^{++}$	&	$^7P_4$		&		 &		 &	0.0033	&	-0.038	&	4.002	 \\

\hline\hline

\end{tabular}

\end{center}
\end{table*}

%%%%%%%%%%%%%%%%%%%%%%%%%%%%%%%%%%%%%%%%%%%%%%%%%%%%%%%%%%%%%%%%%%%%%%%%%%%%%%%%%%%%%%%%%%%%%%%%%%%%%%%%%%%%%%

\begin{table*}
\begin{center}
\caption{Mass spectra of four quark state with $Q\bar q-\bar Qq$ molecular-like structure for $L_1$=1 and $L_2$=1 (in GeV).} \label{tab6}
\begin{tabular}{cccccccccccccccc}
\hline\hline\\ \\

$S_1$	&	$L_1$	&	$S_2$	&	$L_2$&	$J_1$	&	$J_2$	&	$J_{12}$	&	$L_{12}$	&	J	&	 $J^{PC}$  	&	$^{2S+1}X_J$	&	 $M_{cw}$	&	 $\langle V_{SS}\rangle$	&	$\langle V_{LS}\rangle$	&	 $\langle V_{T}\rangle$	&	Mass	\\
\hline\\

0	&	1	&	0	&	1	&	1	&	1	&	0	&		&	0	&	$0^{- +}$	&	$^1S_0$	&		&	 -0.076	&	0	&	0	&	3.960	\\
	&		&		&		&		&		&	1	&	0	&	1	&	$1^{--}$	&	$^3S_1$	&	4.037	 &	 -0.038	&	0	&	0	&	3.998	\\
	&		&		&		&		&		&	2	&		&	2	&	$2^{-+}$	&	$^5S_2$	&		&	 0.038	&	0	&	0	&	4.075	\\
	&		&		&		&		&		&		&		&		&		&		&		&		&		&		 &		\\
	&		&		&		&		&		&	0	&	1	&	1	&	$1^{+-}$	&	$^1P_1$	&	4.143	 &	 -0.0010	&	0	&	-0.0093	&	 4.133	\\
	&		&		&		&		&		&		&		&		&		&		&		&		&		&		 &		\\
	&		&		&		&		&		&	1	&	1	&	0	&	$0^{++}$	&	$^3P_0$	&		&		 &	-0.0022	&	-0.018	&	4.122	\\
	&		&		&		&		&		&		&		&	1	&	$1^{++}$	&	$^3P_0$	&	4.143	 &	 -0.0005	&	-0.0011	&	-0.0046	&	 4.137	 \\
	&		&		&		&		&		&		&		&	2	&	$2^{++}$	&	$^3P_2$	&		&		 &	0.0011	&	-0.010	&	4.134	\\
	&		&		&		&		&		&		&		&		&		&		&		&		&		&		 &		\\
	&		&		&		&		&		&	2	&	1	&	1	&	$1^{+-}$	&	$^5P_1$	&		&		 &	-0.0033	&	-0.029	&	4.112	\\
	&		&		&		&		&		&		&		&	2	&	$2^{+-}$	&	$^5P_2$	&	4.143	 &	 0.0005	&	-0.0011	&	0.010	&	 4.153	\\
	&		&		&		&		&		&		&		&	3	&	$3^{+-}$	&	$^5P_3$	&		&		 &	0.0012	&	-0.014	&	4.131	\\
\hline\\
1	&	1	&	0	&	1	&	0	&	1	&	1	&	0	&	1	&	$1^{--}$	&	$^3S_1$	&	4.037	 &	 0	&	0	&	0	&	4.037	\\
	&		&		&		&		&		&		&		&		&		&		&		&		&		&		 &		\\
	&		&		&		&		&		&	0	&		&	0	&	$0^{-+}$	&	$^1S_0$	&		&	 -0.076	&	0	&	0	&	3.960	\\
	&		&		&		&	1	&	1	&	1	&	0	&	1	&	$1^{--}$	&	$^3S_1$	&	4.037	 &	 -0.038	&	0	&	0	&	3.998	\\
	&		&		&		&		&		&	2	&		&	2	&	$2^{-+}$	&	$^5S_2$	&		&	 0.038	&	0	&	0	&	4.075	\\
	&		&		&		&		&		&		&		&		&		&		&		&		&		&		 &		\\
	&		&		&		&		&		&	1	&		&	1	&	$1^{--}$	&	$^3S_1$	&		&	 -0.1149	&	0	&	0	&	3.922	\\
	&		&		&		&	2	&	1	&	2	&	0	&	2	&	$2^{-+}$	&	$^5S_2$	&	4.037	 &	 -0.038	&	0	&	0	&	3.998	\\
	&		&		&		&		&		&	3	&		&	3	&	$3^{--}$	&	$^7S_3$	&		&	 0.076	&	0	&	0	&	4.113	\\
	&		&		&		&		&		&		&		&		&		&		&		&		&		&		 &		\\

\hline\hline

\end{tabular}

\end{center}
\end{table*}

%%%%%%%%%%%%%%%%%%%%%%%%%%%%%%%%%%%%%%%%%%%%%%%%%%%%%%%%%%%%%%%%%%%%%%%%%%%%%%%%%%%%%%%%%%%%%%%%%%%%%%%%%%%%%%%%%%
\begin{table*}
\begin{center}
\caption{Comparison of some predicted states with experimental results (in GeV)}. \label{tab7}
\begin{tabular}{ccccccccc}
\hline\hline
&   & \multicolumn{2}{c}{Present}               & &  &\multicolumn{2}{c}{Experiment} &\\
\cline{3-4}\cline{7-9}\\
&	State	&	Mass	&	$J^{PC}$	&	$^{2S+1}X_J$	&	Structure of 	&	Mass	&	$J^{PC}$	\\
&           &           &               &                   & four quark state  &           &               \\
\hline

&		&		&		&		&		&		&		\\
&	$X(3823)$	&	3.823	&	$1^{--}$	&	$^3S_1$	&	$Q\bar q-\bar Qq$ 	&$3.823\pm0.0019$\cite{Bhar}&	 $?^{?-}$		 &		\\ \\
&	$Z_c(3885)$	&	3.882	&	$1^{+-}$	&	$^3S_1$	&	$Qq-\bar Q\bar q$ 	 &$3.883^{\pm0.0015}_{\pm0.0042}$\cite{PRL}&	 $1^{+?}$		 &		\\
&		&		&		&		&		&		&		\\
&	$Z_c(3900)$	&	3.897	&	$1^{+-}$	&	$^5P_1$	&	$Q\bar q-\bar Qq$ 	 &$3.899^{\pm0.0036}_{\pm0.0049}$\cite{liu,M Abl}&	 $?^{?-}$		&		 \\
&		&		&		&		&		&		&		\\
&	$X(3915)$	&	3.917	&	$3^{+-}$	&	$^5P_3$	&	$Q\bar q-\bar Qq$ 	 &$3.915^{\pm0.003}_{\pm0.002}$\cite{SU}		&	 ${0/2}^{?+}$	 \\
&		&	3.919	&	$2^{++}$	&	$^3P_2$	&	$Q\bar q-\bar Qq$ 	&		&		\\
&		&	3.915	&	$0^{++}$	&	$^3P_0$	&	$Q\bar q-\bar Qq$	&		&		\\
&		&	3.910	&	$1^{+-}$	&	$^3S_1$	&	$Qq-\bar Q\bar q$ 	&		&		\\
&		&		&		&		&		&		&		\\

&	$X(3940)$	&	3.946	&	$2^{++}$	&	$^5S_2$	&	$Qq-\bar Q\bar q$ 	&	 $3.942^{+0.009}_{-0.008}$\cite{P}	&	 $?^{?+}$	 \\
&		&	3.935	&	$1^{+-}$	&	$^1P_1$	&	$Q\bar q-\bar Qq$ 	&	 	&	 	 \\
&		&	3.939	&	$2^{+-}$	&	$^5P_2$	&	$Q\bar q-\bar Qq$ 	&		&		\\
&		&		&		&		&		&		&		\\
&	$Y(4008)$	&	3.998	&	$1^{- -}$	&	$^3S_1$	&	$Q\bar q-\bar Qq$ 	&	 $4.008^{+0.121}_{-0.049}$\cite{CZ}	&	 $1^{--}$	 \\
&		&	3.997	&	$1^{- -}$	&	$^3D_1$	&	$Q\bar q-\bar Qq$ 	&		&		\\

&		&	3.988	&	$1^{- -}$	&	$^5P_1$	&	$Q\bar q-\bar Qq$ 	&		&		\\
&		&		&		&		&		&		&		\\
&	$Z_c
(4025)$	&	4.023	&	$0^{++}$	&	$^3P_0$	&	$Q\bar q-\bar Qq$ 	&	$4.026^{\pm0.0026}_{\pm0.0037}$	 \cite{M Abl1}&	 $1^{+?}$	 \\
&		&	4.026	&	$1^{+-}$	&	$^1P_1$	&	$Q\bar q-\bar Qq$ 	&		&		\\
&		&	4.027	&	$2^{++}$	&	$^3P_2$	&	$Q\bar q-\bar Qq$ 	&		&		\\
&		&		&		&		&		&		&		\\
&	$\psi(4040)$	&	4.037	&	$1^{- -}$	&	$^3S_1$	&	$Q\bar q-\bar Qq$ 	&	$4.039\pm0.001$	 \cite{Beringer}&	 $1^{- -}$		\\
&		&	4.038	&	$1^{+ +}$	&	$^3P_1$	&	$Q\bar q-\bar Qq$ 	&		&		\\
&		&		&		&		&		&		&		\\

&	$Z_1(4050)$		&	4.046	&	$2^{+ -}$	&	$^5P_2$	&	$Q\bar q-\bar Qq$	 &$4.051^{+0.024}_{-0.043}$\cite{Mizuk,JP}		&	$?^{??}$	\\
&		&	4.054	&	$3^{++}$	&	$^7P_3$	&	$Q\bar q-\bar Qq$ 	&		&		\\
&		&		&		&		&		&		&		\\
&	$Y(4140)$	&	4.136	&	$0^{-+}$	&	$^3P_0$	&	$Qq-\bar Q\bar q$ 	&		&		\\
&		&	4.142	&	$3^{--}$	&	$^5P_3$	&	$Qq-\bar Q\bar q$ 	&	$4.143\pm0.003$	\cite{T1}&	$?^{?+}$	 \\
&		&	4.145	&	$1^{--}$	&	$^1P_1$	&	$Qq-\bar Q\bar q$ 	&		&		\\
&		&	4.146	&	$2^{-+}$	&	$^3P_2$	&	$Qq-\bar Q\bar q$ 	&		&		\\
&		&	4.137	&	$1^{++}$	&	$^3P_1$	&	$Q\bar q-\bar Qq$	&		&		\\
&		&		&		&		&		&		&		\\
&	$X(4160)$	&	4.153	&$2^{+-}$	&	$^5P_2$	&	$Q\bar q-\bar Qq$ 	&	 $4.156^{+0.029}_{-0.025}$\cite{P}	&	 $?^{?+}$	 \\
&			&4.164		&$1^{--}$	&	$^1P_1$	&	$Qq-\bar Q\bar q$ 		&	&     &		\\
&			&4.159		&$1^{-+}$	&	$^3P_1$	&	$Qq-\bar Q\bar q$ 	&	&     &			\\

\hline\hline

\end{tabular}
\begin{center}
$[Q\bar q-\bar Qq]$ - molecular-like structure.\\
$[Qq-\bar Q\bar q]$ - diquark-antidiquark structure.\\

\end{center}
\end{center}

\end{table*}
%%%%%%%%%%%%%%%%%%%%%%%%%%%%%%%%%%%%%%%%%%%%%%%%%%%%%%%%%%%%%%%%%%%%%%%%%%%%%

%%%%%%%%%%%%%%%%%%%%%%%%%%%%%%%%%%%%%%%%%%%%%%%%%%%%%%%%%%%%%%%%%%%%%%%%%%%%%%%%%%%%%%%%%%%%%%%%
\begin{table*}
\begin{center}
\caption{The ground state mass and $1^{st}$ radial excitation state mass of predicted states in Table 7 (in GeV). We have compared $1^{st}$ radilly excited mass with available experimental data.}\label{tab8}
\begin{tabular}{ccccccc}
\hline\hline\\

State	&	$J^{PC}$	&	Model	&	$M_{ground}$	&	$M_{rad. excitation}$	&	 Experiment
 	\\
%        &               &           &                   &                   &    excited state  \\
\hline\\
$X(3823)$	&	$1^{--}$	&	$Q\bar q-\bar Qq$	&	3.823	&	4.247	&		\\
	&		&		&		&		&		\\
$Zc(3885)$	&	$1^{+-}$	 &	$Qq-\bar Q\bar q$	&	3.882	&	4.426	 &[$Z(4430)$]$4.475{\pm0.007}^{+0.015}_{-0.025}$\cite{RA}		\\
	&		&		&		&		&		\\
$Zc(3900)$	&	$1^{+-}$	&	$Q\bar q-\bar Qq$	&	3.897	&	4.312	&		\\
	&		&		&		&		&		\\
$X(3915)$	&	$3^{+-}$	&	$Q\bar q-\bar Qq$	&	3.917	&	4.328	&		\\
	&	$2^{++}$	&	$Q\bar q-\bar Qq$	&	3.919	&	4.330	&		\\
	&	$0^{++}$	&$Q\bar q-\bar Qq$	&	3.915	&	4.388	&		\\
	&	$1^{+-}$	&	$Qq-\bar Q\bar q$	&	3.910	&	4.327	&		\\
	&		&		&		&		&		\\
$X(3940)$	&	$2^{++}$	&	$Qq-\bar Q\bar q$	&	3.946	&	4.408	&		\\
	&	$1^{+-}$	&	$Q\bar q-\bar Qq$	&	3.935	&	4.343	&		\\
	&	$2^{+-}$	&$Q\bar q-\bar Qq$	&	3.939	&	4.346	&		\\ \\
$Y(4008)$	&	$1^{--}$	&	$Q\bar q-\bar Qq$	&	3.998	&	4.400	&		\\
	&	$1^{--}$	&	$Q\bar q-\bar Qq$	&	3.997	&	4.389	&		\\
	&	$1^{--}$	&	$Q\bar q-\bar Qq$	&	3.988	&	4.360	&	 [$Y(4360)$]$4.361^{\pm0.009}_{\pm0.009}$\cite{XL}	\\ \\
$Zc(4025)$	&	$0^{++}$	&	$Q\bar q-\bar Qq$	&	4.023	&	4.400	&		\\
	&	$1^{+-}$	&	$Q\bar q-\bar Qq$	&	4.026	&	4.402	&		\\
	&	$2^{++}$	&	$Q\bar q-\bar Qq$	&	4.027	&	4.403	&		\\
	&		&		&		&		&		\\
$\psi(4040)$	&	$1^{--}$	&	$Q\bar q-\bar Qq$	&	4.037	&	4.413 &[$\psi(4415)$] $4.421\pm0.004$	 \cite{Beringer}		\\
	&	$1^{++}$	&	$Q\bar q-\bar Qq$	&	4.038	&	4.413	&		\\
	&		&		&		&		&		\\
$Z_1(4050)$	&	$2^{-+}$ 	&	$Q\bar q-\bar Qq$	&	4.046	&	4.419	&		\\
	&	$3^{++}$	&	$Q\bar q-\bar Qq$	&	4.0514	&	4.426	&		\\
	&	$3^{--}$	&	$Qq-\bar Q\bar q$	&	4.142	&	4.535	&		\\ \\
$Y(4140)$	&	$0^{-+}$	&	$Qq-\bar Q\bar q$	&	4.136	&	4.530	&		\\
	&   $1^{--}$	&	$Qq-\bar Q\bar q$	&	4.145	&	4.537	&		\\
	&	$2^{-+}$	&	$Qq-\bar Q\bar q$	&	4.146	&	4.538	&		\\
	&	$1^{++}$	&	$Q\bar q-\bar Qq$	&	4.137	&	4.479	&		\\
	&		&		&		&		&		\\
$X(4160)$	&	$2^{+-}$	&	$Q\bar q-\bar Qq$	&	4.153	&	4.492	&		\\
	&	$1^{-+}$	&	$Qq-\bar Q\bar q$	&	4.159	&	4.548	&		\\
	&	$1^{--}$	&	$Qq-\bar Q\bar q$	&	4.164	&	4.549	&		\\

\hline\hline

\end{tabular}

\end{center}
\end{table*}
%%%%%%%%%%%%%%%%%%%%%%%%%%%%%%%%%%%%%%%%%%%%%%%%%%%%%%%%%%%%%%%%%%%%%%%%%%%%%%%%%%%%%%%%%%%%%%%%

\subsection{The Four quark state as $Qq$, $\bar Q\bar q$ clusters}
In this section, we calculate the mass spectra of tetraquarks with hidden charm as the bound states of two clusters ($Qq$ and $\bar Q\bar q$), ($Q=c ;\ \ q=u,d$). We think of the diquarks as two correlated quark with no internal spatial excitation. Because a pair of quarks can't be a color singlet, the diquark can only be found confined into the hadrons and used as effective degree of freedom. Heavy light diquarks can be the building blocks of a rich spectrum of exotic states which can not be fitted in the conventional charmonium assignment. Maiani et al \cite{Maiani} in the framework of the phenomenological constituent quark model considered the masses of hidden/open charm diquark-antidiquark states in terms of the constituent diquark masses with their spin-spin interactions included.

We discuss the spectra in the framework of a non-relativistic hamiltonian including chromo-magnetic spin-spin interactions between the quarks (antiquarks) within a diquark(antidiquark ). Masses of diquark (antidiquark) states are obtained by numerically solving the  Schr\"{o}dinger equation with the respective two body potential given by Eq.(\ref{eq:9}) and incorporating the respective spin interactions described by equation (\ref{eq:10}) perturbatively.\\
 In the diquark-antidiquark  structure, the masses of the  diquark/diantiquark system are given by:

 \begin{equation}\label{eq:14}
m_d=m_{Q}+m_{q}+E_{d}+{\langle V_{SD}\rangle_{Qq}}
\end{equation}

\begin{equation}\label{eq:15}
m_{\bar d}=m_{\bar Q}+m_{\bar q}+E_{\bar d}+{\langle V_{SD}\rangle_{\bar Q\bar q}}
\end{equation}
Further, the same procedure is adopted to compute the binding energy of the diquark-antidiquark  bound system as
\begin{equation}\label{eq:16}
M_{d-\bar d}=m_d+m_{\bar d}+E_{d\bar d}+{\langle V_{SD}\rangle}_{d\bar d}
\end{equation}

Where $Q$ and $q$ represents the heavy quark and light quark respectively. Here, we consider a different string tension $\sigma_{Qq-\bar Q\bar q}$ for intercluster interaction. In the present paper, $d$ and $\bar d$ represents diquark and antidiquark respectively. While $E_d$, $E_{\bar d}$, $E_{d \bar d}$ are the energy eigen values of the diquark, antidiquark and diquark-antidiquark system respectively. The spin-dependent potential ($V_{SD}$) part of the hamiltonian described by Eq.(\ref{eq:10}) has been treated perturbatively. Details of the computed results are listed in Table 2 and 3 for the low lying positive parity and negative parity states respectively.\\

\subsection{Four quark state as $Q\bar q$, $\bar Qq$ clusters}

  In this section, we present our calculations for the masses of the four quark states as clusters of $Q\bar q$ and $\bar Qq$  with some residual color forces that binds the two clusters. Unlike in the di-mesonic molecular bound states where screening potentials are employed \cite{Wong} for the interaction between two color singlet mesons, here we consider a residual color interaction among the $Q\bar q$, $\bar Qq$ clusters within the four quark system but with a screened string tension ${\sigma_{Q\bar q-\bar Qq}}$. Accordingly we write ${\sigma_{Q\bar q-\bar Qq}}=e^{-x}{\sigma_{Qq-\bar Q\bar q }}$, where the screening factor $e^{-x}$ is found to be about $60\%$. Apart from this, we have also considered one meson exchange interaction at the long range part of the $Q\bar q-\bar Qq$ interaction with the usual meson-nucleon coupling strength and we found that the contributions are very negligible compared to their spin dependent contributions.\\ \\
  The masses of two clusters ($Q\bar q$, $\bar Qq$) are obtained by solving the  Schr\"{o}dinger equation with the intra-cluster interactions of the Coulomb plus linear form given as follows:
 \begin{equation}\label{eq:7}
M_1=m_{Q}+m_{\bar {q}}+E_{Q\bar {q}}
\end{equation}

\begin{equation}\label{eq:7}
M_2=m_{\bar Q}+m_{ q}+E_{\bar{ Q}q}
\end{equation}\\

Further, the calculations are repeated for the mass of the four quark system incorporating the intercluster interaction and is expressed as

\begin{equation}\label{eq:7}
M=M_1+M_2+E_{M_{12}}+\langle V_{SD}\rangle
\end{equation}
where, $E_{Q\bar {q}}$, $E_{\bar{ Q}q}$ represent the binding energy of the quark-antiquark constituting the cluster and $E_{M_{12}}$ is the effective binding energy among the two clusters. The interaction between the two quark-antiquark systems are also assumed to be of the same form as given by Eq.(\ref{eq:9}) except that the string tension is assumed to be different. And the term ($V_{SD}$) represents the spin-dependent potential part of two body cluster($Q\bar q-\bar Qq$) given by  Eq.(\ref{eq:10}).

 Here, we have taken various combinations of spin and orbital angular momentum. We have considered the total spin $J_1$ and $J_2$ of the two quark-antiquark clusters as spins $S_1$ and $S_2$ and these spins couple to $J_{12}$ together with relative orbital motion $L_{12}$ presents the total spin J of the $Q\bar q-\bar Qq$  system. The computed masses of the four quark system with different combinations of the spin and orbital angular momenta are given in Tables 4, 5 and 6.  \\

%%%%%%%%%%%%%%%%%%%%%%%%%%%%%%%%%%%%%%%%%%%%%%%%%%%%%%
\section{Results and Discussions}

 The masses of the low lying hidden charm four quarks states as $Qq-\bar Q\bar q$ clusters as well as $Q\bar q-\bar Qq$ clusters have been computed. Various combinations of the orbital and spin excitations have been considered. The results obtained in both the cases are tabulated in Tables 2 to 6. Selected states for known experimental exotic states are identified and their $J^{PC}$ values are assigned. Their interpretations are shown in Table 7. The first radial excitations of these predicted states are also computed and the results along with their ground state mass are listed in Table 8. Finally, we find it interesting to compare our results with the newly discovered exotic charm states. For example, soon after the $ Z^+_{c}(3900)$ observation, the BES-III reported the observation of three more charge states: $Z^+_{c}(4025)$ \cite{M Abl1}, $Z^+_{c}(4020)$ \cite{M Abl2} and $Z^+_{c}(3885)$ \cite{PRL}. \\Here, we have been able to identify the  $X(3823)$, $Z_c(3900)$, $X(3915)$, $Z_c(4025)$, $X(4160)$ and  $\psi(4040)$  resonances as $Q\bar q-Q\bar q$ clusters of molecular-like states, while $Z_c(3885)$, $X(3940)$, $Y(4140)$ as diquark-antidiquark cluster states. The charge conjugation of $X(3823)$ and $Z_c(3900)$ states is reported by BELLE and BES-III as C=-1, but their $J^P$ values are not assigned experimentally \cite{liu,M Abl,Bhar}. Our predictions suggest $Z_c(3900)$ as $1^{+-}$ state while $X(3823)$ state as $1^{--}$. There is a still question regarding the structure of two states $Z_c(3900)$ and $Z_c(3885)$ that whether they are two different states or the same state. Recently, BES-III group reported that $Z_c(3885)$ may have $1^+$ quantum number and if so then it can be a S wave or/and a D wave state \cite{PRL}. Our present study predicts $Z_c(3885)$ as ($^3S_1$) having diquark-antidiquark  structure with $J^{PC}$ =$1^{+-}$. Although the $J^P$ quantum numbers of $Z_c(4025)$  still remain to be determined experimentally, it is assumed to have spin parity $J^P = 1^+$  by BES-III group\cite{M Abl1}. This assignment for $Z_c(4025)$ is in agreement with the interpretation of this state to be $Q\bar q-\bar Qq$ molecular like state having $J^{PC} = 1^{+-}$. Present identification of $\psi(4040)$ as a $Q\bar q-\bar Qq$ molecular like state with $J^{PC} = 1^{--}$ is in accordance with what was suggested by De Rujula, Georgi and Glashow \cite{Rujula}. The $J^{PC}$ value of $X(3940)$ state is still not confirmed.
 It can be interpreted as either $1^{+-}/2^{+-}$ $Q\bar q-\bar Qq$  molecular-like state or a $2^{++}$ diquark-antidiquark state. In the present calculation, our X(3940) state as a $2^{++}$ diquark-antidiquark state is in good agreement with Maiani et al\cite{Maiani} and compatible with experiment. In Ref \cite{Maiani}, authors have also predicted X(3940) state as a $2^{++}$ diquark-antidiquark state. The state $Z_{1}(4050)$ is close to the  interpretations of $Q\bar q-\bar Qq$  molecular like state having same positive parity
 but with different J values. Thus, state $Z_1(4050)$ still needs more experimental confirmation
 for its J value. From our present study, we assign $J^{PC}=2^{+-}$ or $3^{++}$ if it is having  $Q\bar q-\bar Qq$ molecular-like structure. Up to now, the interpretation of the state $X(4160)$ and  $X(3915)$ is still unclear. The state $X(3915)$ clearly has C=+, but $J^P$ remains to be determined. T Branz et al. \cite{T} predicted this state as a molecular state. But we have four possibilities for the interpretations of the $X(3915)$ state: it can be either one of the $0^{++}/2^{++}/3^{+-}$ $Q\bar q-\bar Qq$ molecular-like state or $1^{+-}$ diquark-antidiquark state. If we follow experimental clue for C=+, then it could be a $0^{++}/2^{++}$ $Q\bar q-\bar Q q$ molecular-like state. The interpretation of this $X(3915)$ state as $0^{++}$ molecular like state is in agreement with Youchang et al \cite{Youchang}. For $X(4160)$, we have predicted that it can be a either $2^{+-}$
    $Q\bar q-\bar Qq$  molecular-like state or $1^{+-}/1^{-+}$ diquark-antidiquark  state. \\For $Y(4140)$, we are having four different possible states $0^{-+}$, $1^{--}$, $2^{-+}$, $3^{--}$ in the energy range $4.136-4.146$ GeV as diquark -antidiquark  states while only $1^{++}$ state in the $Q\bar q-\bar Q q$  molecular-like structure. As per C=+1 assignment provided by the experiment \cite{T1}, then it can be
     interpreted as a $Q\bar q-\bar Qq$  molecular-like state only if its parity is positive. However, its experimental confirmation is awaited. In Ref. \cite{LFAD}, authors have re-examined the four quark  system in diquark-antidiquark model especially in $J^{PG}=1^{++}$ and $1^{--}$ sectors. They have interpreted $Y(4360)$ as the first radial excitation of $Y(4008)$  with $J^{PG} =1^{--}$. They found that mass difference between these two states is 350 $MeV$ which is very similar to that of mass difference of ground and radial excitation of P wave bottomonia.
In our present calculation, we also interpreted $Y(4360)$ with $J^{PC} =1^{--}$ (negative G parity) as  first radial excited state of $Y(4008)$ in molecular picture. We have found relative mass difference of 362 $MeV$ in the range of mass difference for L=1 bottomonia i.e. $\chi_{bJ}(2P)- \chi_{bJ}(1P)=360MeV$ in accordance with Maiani et al.
 In Ref.\cite{LFAD} authors interpreted  $Z(4430)$  state  as  the first radial excitation of the $Z(3900)$, with a mass difference $Z(4430)-Z(3900) = 593 MeV$, very close to  $\psi(2S)-\psi(1S) =589 MeV$.
Our present study suggests that $\psi(4415)$ can be the first radial excitation of the $\psi(4040)$ in the $Q\bar q-\bar Qq$  molecular-like structure having $J^{PC}=1^{--}$. In our present calculation , we found that  $Z(4430)$  can be the first radial excitation state of the $Zc(3885)$($J^{PC} =1^{+-}$ with positive G parity) in diquark-antidiquark picture with a mass difference of 543 $MeV$ which is lower than that reported by Ref. \cite{LFAD}.\\
The present study has been able to identify many newly observed exotic states with their $J^{PC}$ assignment. Finally, We believe that more experimental efforts aimed at determining the spin parity of the exotic states are required for understanding the structure of many of the newly observed exotic states. We also hope that many of the predicted exotic states will be supported by the future experimental observations.

\section{Acknowledgments}%\acknowledgments
The work is part of Major research project NO. F. 40-457/2011(SR) funded by UGC, INDIA.

%%%%%%%%%%%%%%%%%%%%%%%%%%%%%%%%%%%%%%%%%%%%%%%%%


\begin{thebibliography}{}
% and use \bibitem to create references.

\bibitem{liu} Z. Q. Liu et al. (Belle Collaboration), Phys. Rev. Lett. {\bf110}, 252002 (2013).
\bibitem{M Abl} M. Ablikim et al.,(The BESIII Collaboration), Phys. Rev. Lett. {\bf110}, 252001 (2013).
\bibitem{JW} R. L. Jaffe, F. Willczek, Phys. Rev. Lett. {\bf91}, 232003 (2003).
\bibitem{Jaffe} R. L. Jaffe, Phys. Rev. D {\bf15}, 281 (1977).
\bibitem{R}     R. L. Jaffe, Phys. Rev. D {\bf15}, 267 (1977).
\bibitem{hooft} G. 't Hooft, G. Isidori, L. Maiani, A.D. Polosa, V. Riquer, Physics Lett. B {\bf662}, 424(2008).
\bibitem{Lmaiani} L. Maiani, F. Piccinini, A.D. Polosa, V. Riquer, Phys. Rev. Lett.{\bf93}, 212002 (2004).
\bibitem{black} D. Black, A. Fariborz, J. Schechter, Phys. Rev. D {\bf61} ,074001(2000).

\bibitem{W} J. D. Weinstein and N. Isgur, Phys. Rev. D {\bf27}, 588 (1983).
\bibitem{Isgur} J. D. Weinstein and N. Isgur, Phys. Rev. D {\bf41}, 2236 (1990).
\bibitem{Voloshin}M.B. Voloshin, L.B. Okun, JETP Lett. {\bf23}, 333 (1976).
\bibitem{Youchang}Youchang Yanga and Jialun Ping, arXiv:1004.2444.
\bibitem{Wong} Cheuk-Yin Wong, Phys.Rev. C {\bf79}, 055202(2004).
\bibitem{Rui}Rui Zhang, Yi-Bing Ding, Xue-Qian Li, and Philip R. Page, Phys. Rev. D {\bf65}, 096005 (2002).
\bibitem{Choi}S. K. Choi et al. [BELLE Collaboration],  Phys. Rev. Lett. {\bf100}, 142001 (2008).

\bibitem{RA}R. Aaij et al. [LHCb Collaboration], Phys. Rev. Lett. {\bf112}, 222002 (2014).

%\bibitem{N}N. A. Tornqvist, Z. Phys. C {\bf61}, 525 (1994).
\bibitem{Dias} J. M. Dias,  F. S. Navarra, and M. Nielsen, Phys. Rev. D {\bf88}, 016004 (2013).
\bibitem{LM}L. Maiani, V. Riquer, R. Faccini, F. Piccinini, A. Pilloni and A. D. Polosa, Phy. Rev. D {\bf87}, 111102(R) (2013).
\bibitem{Hong}Hong-Wei Ke et al., Eur. Phys. J. C {\bf73}, 2561 (2013).
\bibitem{Lucha} Lucha W and Shoberl F, Int. J. Mod. Phys. C {\bf10}, (1999),  arXiv:hep-ph[9811453].
\bibitem{BKP} Bhavin Patel and P C Vinodkumar, J Phys. G: Nucl. Part. Phys. {\bf36}, 035003, (2009).
\bibitem{Maiani}L. Maiani, F. Piccinini, A. D. Polosa and V. Riquer, Phys. Rev. D {\bf71}, 014028 (2005) [hep-ph/0412098].

%\bibitem{jafferep}R. L. Jaffe, Phys. Rep. {\bf409}, 1 (2005).
%\bibitem{jaffehep}R. L. Jaffe (1999), hep-ph/0001123.
%\bibitem{Lichtenberg}D. B. Lichtenberg, R. Roncaglia, and E. Predazzi (1996),hep-ph/9611428.

\bibitem{M Abl1}M. Ablikim et al. [BESIII Collaboration], arXiv:1308.2760.
\bibitem{M Abl2}M. Ablikim et al. [BESIII Collaboration], arXiv:1309.1896.
\bibitem{PRL}M. Ablikim et al. [BESIII Collaboration], Physics Rev. Lett. {\bf112}, 022001(2014).
\bibitem{Bhar} V. Bhardwaj et al. [Belle Collaboration], accepted for publication in Phys. Rev. Lett,
               arXiv:1304.3975 [hep-ex].
\bibitem{Rujula} A. De Rujula, H. Georgi, S.L. Glashow, Phys. Rev. Lett. {\bf38}, 317 (1977).

\bibitem{SU} S. Uehara et al. [Belle Collaboration], Phys. Rev. Lett. {\bf104}, 092001 (2010).


\bibitem{P} P. Pakhlov et al. (Belle Collaboration), Phys. Rev. Lett. {\bf100}, 202001 (2008).
\bibitem{CZ}C. Z. Yuan et al. (Belle Collaboration), Phys. Rev. Lett. {\bf99}, 182004 (2007).
\bibitem{Beringer} J. Beringer et al. (Particle Data Group), Phys. Rev. D {\bf86}, 010001 (2012).
\bibitem{Mizuk} R. Mizuk et al. (Belle Collaboration), Phys. Rev. D {\bf78},  072004 (2008).
\bibitem{JP} J. P. Lees et al. (BABAR Collaboration), arXiv:1111.5919 [hep-ex].
\bibitem{T1} T. Aaltonen et al. (CDF Collaboration), Phys. Rev. Lett. {\bf102}, 242002 (2009).
\bibitem{T} Tanja Branz, Thomas Gutsche and Valery E. Lyubovitskij, Phys. Rev D {\bf80}, 054019 (2009).
\bibitem{XL}X. L. Wang et al. [Belle Collaboration], Phys. Rev. Lett. {\bf99}, 142002
(2007), arXiv:0707.3699v2 [hep-ex].
\bibitem{LFAD}L. Maiani, F. Piccinini, A. D. Polosa. and V. Riquer, accepted for publication in Phys. Rev. D,
              arXiv:1405.1551v2 [hep-ph].
%\bibitem{Vanroyenaweissskopf} R Van Royen and V F Weisskopf, Nuovo Cimento 50 (1967).
%\bibitem{Hwang1997} D S Hwang and Gwang-Hee Kim, Z. Phys. C 76 (1997) 107.
%\bibitem{Bhavin2009} B Patel and P C Vinodkumar, J. Phy. G : Nucl Part. Phy. 36 (2009) 035003.
%\bibitem{Bodwin1995} G T Bodwin, J Lee and D K Sinclair, Phys. Rev. D51 (1995) 1125.
%\bibitem{Berezhnoy1996} A V Berezhnoy, V V Kiselev and A K Likhoded, Z. Physik A 336 (1996) 89 .
%\bibitem{EBraaten1995} E Braaten and S Fleming, Phys. Rev. D 52 (1995) 181.
%\bibitem{Gerstein1998} S S Gerstein et al., Int. J. Mod. Phys. A 6 (1991) 2309.


\end{thebibliography}
\end{document}